# Magnetic non-uniformity and thermal hysteresis of magnetism in a manganite thin film

Surendra Singh[1,2], M. R. Fitzsimmons[1], T. Lookman[1], J. D. Thompson[1], H. Jeen[3,4], A. Biswas[3], M. A. Roldan[5] and M. Varela[4]

[1]Los Alamos National Laboratory, Los Alamos, NM 87545, USA

[2] Solid State Physics Division, Bhabha Atomic Research Center, Mumbai 400085, India

[3]Department of Physics, University of Florida, Gainesville, FL 32611, USA

[4]Oak Ridge National Laboratory, Oak Ridge TN 37831 USA

[5]University Complutense, Madrid 28040, Spain.

**Abstract:** We measured the chemical and magnetic depth profiles of a single crystalline $(La_{1-x}Pr_x)_{1-y}Ca_yMnO_{3-\delta}$ ($x = 0.52\pm0.05$, $y = 0.23\pm0.04$, $\delta = 0.14\pm0.10$) film grown on a $NdGaO_3$ substrate using x-ray reflectometry, electron microscopy, electron energy-loss spectroscopy and polarized neutron reflectometry. Our data indicate that the film exhibits coexistence of different magnetic phases as a function of depth. The magnetic depth profile is correlated with a variation of chemical composition with depth. The thermal hysteresis of ferromagnetic order in the film suggests a first order ferromagnetic transition at low temperatures.

Doped bulk perovskite manganites, such as $(La_{1-x}Pr_x)_{1-y}Ca_yMnO_3$, exhibit a rich variety of electronic, magnetic, structural phenomena and phases, which are closely coupled to atomic structure and strain [1-4]. This coupling produces interesting non-linear responses to the environment [4] including, colossal magnetoresistance (CMR) and colossal elastoresistance [5-7]. Considerable theoretical and experimental studies on bulk $(La_{1-x}Pr_x)_{1-y}Ca_yMnO_3$ suggest collective charge-spin-orbital-lattice interactions lead to coexistence of different phases forming domains, viz. cubic ferromagnetic metallic, orthorhombic antiferromagnetic charge ordered insulating, and pseudo-cubic paramagnetic insulating phases. Dimensions of the minority phase domains range from nanometers to microns [3, 8-10]. The competition between the different phases is crucial to understanding macroscopic properties such as CMR and the metal-insulator-transition (MIT) in bulk materials. Properties of thin films of doped perovskites are even more difficult to understand than their bulk counterparts. For example, epitaxial strain between the film and substrate adds a further degree of complexity by influencing magnetic ordering and electronic transport [11-12].

Because manganite thin films exhibit non-linear response such as CMR near the MIT [13], they have enormous potential for a variety of applications, for example as magnetic sensors and tunnel junctions. Owing to high spin polarization at the Fermi level, manganites thin films (e.g. $La_{0.7}Sr_{0.3}MnO_3$) should be attractive as spin injectors [14-15]. However, several groups have reported unexpectedly low values of tunneling magetoresistance (TMR) for magnetic tunnel junctions using manganite films [16-17]. Spin resolved photoemission spectroscopy (SPRS) studies attribute the loss of TMR to a degraded interfacial magnetization [18] caused by cation segregation to interfaces, which may change the electronic structure in the interfacial region [19-21]. An electron energy-loss spectroscopy (EELS) study of manganite films confirmed segregation of cations to interfaces [21]. While SPRS and EELS provide information about the depth dependence of the chemical structure (which is not necessarily representative of the entire sample), neither technique measures the magnetization depth profile. X-ray reflectivity (XRR) and polarized neutron reflectivity (PNR) are two nondestructive techniques that provide quantitative measures of the chemical and magnetic depth profiles of films with nanometer resolution [22-26] averaged over the lateral dimensions of the entire sample (typically 100 mm$^2$).

We report electrical transport and PNR measurements, in conjunction with magnetometry, EELS and XRR studies of a single crystalline $(La_{1-x}Pr_x)_{1-y}Ca_yMnO_{3-\delta}$ (x = 0.52±0.05, y = 0.23±0.04, δ = 0.14±0.10) (LPCMO) film epitaxially grown on a (110) NdGaO$_3$ (NGO) substrate. PNR allowed us to monitor the evolution of the depth dependence of magnetism as a function of temperature across the MIT. Analysis of XRR and PNR data show that the degradation of magnetization is correlated with chemical non-uniformity, particularly near interfaces. The chemical non-uniformity was confirmed with EELS.

A 35-nm-thick epitaxial LPCMO thin film with the nominal composition of $(La_{1-x}Pr_x)_{1-y}Ca_yMnO_3$ (x = 0.6, y = 0.33) was grown on orthorhombic NGO substrates by pulsed (KrF) laser (wavelength = 248 nm) deposition (PLD). The substrate temperature was kept at 780 °C, O$_2$ partial pressure was 130 mTorr, laser fluence was about 0.5 J/cm$^2$, and repetition rate was 5 Hz [27].

In order to study the chemical non-uniformity along the depth of the film, we carried out EELS measurements (Figs. 1(a)-(c)) of cross-sectional specimens prepared by conventional methods in an aberration corrected Nion UltraSTEM scanning transmission electron microscope operated at 100 kV and equipped with a Gatan Enfina spectrometer. The O/Mn relative concentration (Fig 1(a)) maps were produced using the O-K and Mn-L$_{2,3}$ edges [28]. The Ca, La and Pr relative concentration (Fig. 1(b)) maps were produced using the Ca-$L_{2,3}$, La-$M_{4,5}$ and Pr-$M_{4,5}$ edges, respectively after background subtraction using a power law fit and integration of the intensity under every absorption edge. Principal component analysis was applied to remove random noise [29]. In the middle of the film (region II in Fig. 1(a)) a slight decrease of the O/Mn ratio (~ 2.86 ±0.10) was observed, suggesting an O deficiency in this region. The average composition of the film $(La_{1-x}Pr_x)_{1-y}Ca_yMnO_{3-\delta}$ ($x = 0.52\pm0.05$, $y = 0.23\pm0.04$, $\delta = 0.14 \pm0.10$) was different than the nominal composition of the laser ablated target material. A significant increase (decrease) in the average La (Ca) concentration with respect to the nominal values was observed. However, the Pr concentration was similar (~ 40%) to the concentration of the target material (Fig. 1(b)). The concentrations of these elements remained relatively constant with depth through the film bulk (region II). However small changes in the concentrations of La, Pr and Ca, and an increase of the O/Mn ratio were observed at the surface and buried (film-substrate) interface (regions I and III in Figs. 1(a-c)). The increase of the O/Mn ratio in region III may be somewhat affected by electron beam broadening due to dechanneling within 1-2 nm of the buried interface.

We estimated the Mn valence along the depth of the film from the EELS data (Fig. 1(c)) using two independent methods: (1) as inferred from the chemical composition [30], and (2) from the ratio of intensities of the Mn-$L_{2,3}$ edges [28]. Both methods show an oxidation state of Mn close to +3 in region II, and an increase of the Mn oxidation state near the surface and buried interface (regions I and III). The increase implies a higher concentration of $Mn^{4+}$ near the surface and buried interface.

Previously, an EELS study of LCMO grown on a $SrTiO_3$ substrate found an increase of Mn oxidation state at the surface and a *decrease* of Mn oxidation state at the buried interface [21].

Ref. [21] attributed the variation of Mn oxidation state to a balance between epitaxial strain and kinetic effects during growth. Here, we have observed an *increase* in the Mn oxidation state for the surface and *especially* the buried interface. In contrast to the LCMO/SrTiO$_3$ system, the epitaxial strain in the LPCMO/NGO system is much smaller [27]. We suggest that the change of Mn oxidation state is related to the variation of cations, especially Pr, across the film and the concomitant change of strain from the small ionic size of Pr compared to La. Regardless of its origin, the variation of Mn valence across the film's depth may affect the magnetization depth profile because the magnetic moments of Mn$^{3+}$ and Mn$^{4+}$ are different.

Macroscopic magnetization measurements were performed using VSM and SQUID magnetometry. Fig. 1(d) shows evidence for magnetic anisotropy in the plane of the film (with easy axis parallel to [1$\bar{1}$0] NGO) at a temperature of 20 K. The electrical transport (resistance) measurements were taken while changing the temperature of the sample at a rate of 0.4 K / min using a two-probe method [6] in a closed cycle helium cryostat during the neutron scattering experiment. Fig. 1(e) shows a comparison of the transport curves measured in zero and 6 kOe fields (applied along easy axis). The sample exhibited a sharp transition in resistance on cooling (insulator to metal, $T_{IM}$) and warming (metal to insulator, $T_{MI}$) cycles with a thermal hysteresis of ~ 17 K at 6 kOe. Magnetic fields stabilize both $T_{IM}$ and $T_{MI}$ to higher temperatures.

The specular reflectivity, *R*, of the sample was measured as a function of wave vector transfer, $Q = 4\pi \sin\theta/\lambda$ (where, $\theta$ is angle of incidence and $\lambda$ is the x-ray or neutron wavelength). The reflectivity is qualitatively related to the Fourier transform of the scattering length density (SLD) depth profile $\rho(z)$ [24, 25] averaged over the whole sample area. For XRR, $\rho_x(z)$, is proportional to electron density [24, 25]. In case of PNR, $\rho(z)$ consists of nuclear and magnetic SLDs such that $\rho^{\pm}(z) = \rho_n(z) \pm CM(z)$, where $C = 2.853\times10^{-9}$ Å$^{-2}$G$^{-1}$, and *M(z)* is the magnetization (in G) depth profile [24]. The +(-) sign denotes neutron beam polarization along (opposite to) the applied field. $\rho_n(z)$ and $M(z)$ can be inferred from $R^{\pm}(Q)$ often with nm+ resolution. The difference between $R^+(Q)$ and $R^-(Q)$ divided by the sum, called the spin asymmetry, $asym = (R^+(Q) - R^-(Q))/(R^+(Q) + R^-(Q))$, can be a very sensitive measure of small $M$. The reflectivity data were normalized to the Fresnel reflectivity ($R_F = \frac{16\pi^2}{Q^4}$) [24].

XRR measurements were carried out using Cu $K_\alpha$ radiation at Los Alamos Neutron Scattering Center (LANSCE). The XRR data (closed circles) shown in the inset of Fig. 2 (a) are dominated by oscillations inversely related to the total thickness of the LPCMO film. However, an additional modulation of the scattering is present, which is evidence for a non-uniform chemical composition across the film's thickness.

PNR measurements (Fig. 2) were carried out using the Asterix spectrometer at LANSCE [24]. The PNR measurements were performed at 6 kOe (applied along easy axis) after cooling the sample at a rate of 0.4 K /min in the same field (6 kOe). $R^{\pm}(Q)$ for 200 K and 20 K are shown in Fig. 2(a). Fig. 2(b) and (c) show the spin asymmetry for the same temperatures, respectively. The open (closed) triangles on transport data in Fig. 2(d) show the temperatures while cooling (warming) the sample across the MIT for which we have also measured $R^{\pm}(Q)$.

The chemical and magnetic density profiles were obtained by fitting a model $\rho(z)$ whose reflectivity best fits the data. The reflectivities were calculated using the dynamical formalism of Parratt [31]. Using the chemical profile from EELS as a guide, we represented the chemical/nuclear depth profile as three layers as shown in Fig. 2(e). This representation produced an acceptable fit to the XRR data (inset of Fig. 2(a)).

We optimized the nuclear SLD profile by constraining layer thicknesses and interface roughness to be within the 95% confidence limit established from the analysis of the XRR data.

We fitted the three-layer-model to PNR data taken well above the Curie temperature (~ 130 K) of the film (at 200 K). The solid (black) curves in Figs 2(a) (for 200K) and (b) were obtained from a calculation of the reflectivity using the nuclear SLD shown as the solid (black) curve in Fig 2(f). Next, the nuclear SLD was fixed and then $M(z)$ was optimized using the PNR data taken at 20 K. The calculated $R^{\pm}(Q)$ are shown by the solid (black and green) curves in Fig 2(a). $M(z)$ is shown as the solid (red) curve in Fig 2(g). The three-layer-model was also fitted to the PNR measurements for the intermediate temperatures (see Fig. 3). The SLD profiles obtained with XRR and PNR suggest non-uniformity chemical composition along the depth of the film. These results are consistent with the EELS study.

The PNR data indicate that the magnetization depth profile is also non-uniform across the depth of the LPCMO film. The variation of the magnetization is a result that can be anticipated from a variation of the Mn valence, i.e., from the change of $Mn^{4+}$ relative to $Mn^{3+}$ [32]. At 20 K, we obtained a magnetization of 635±40 G (4.0±0.3 $\mu_B$ per formula unit) for region II. However, the magnetizations for the surface and buried interface (regions I and III) at 20 K were considerably less, 140±42 and 70±25 G, respectively. These are also the regions in which the stoichiometry of the film is different from the film's bulk. Using the theoretical values of $\mu_{eff}^{th}(Mn^{3+}) = 4.90\mu_B$ and $\mu_{eff}^{th}(Mn^{4+}) = 3.87\mu_B$ and assuming saturation of these moments, we estimate an upper limit on the net moment to be $\mu_{eff}^{cal} = 4.6\ \mu_B$, which is close to that inferred from the magnetization of region II at 20K. Since the moment for $Mn^{4+}$ is smaller than that of $Mn^{3+}$, the decrease of magnetization in the boundary regions (regions I, and III) is consistent with an increased concentration of $Mn^{4+}$ in these regions. Other factors that could lead to suppression of magnetization include: phase separation in the lateral dimensions of the sample, strain, and antiferromagnetic interactions [3]. Our experiment cannot exclude these scenarios.

In Fig 4, we show the magnetization of each region as a function of temperature as obtained from the $M(z)$ profiles shown in Fig. 3. Fig. 4 is remarkable for two reasons. First, we see evidence for thermal hysteresis (12.0±0.3, 14.0±0.3 and 16.0±0.3 K, for regions I, II and III, respectively) of the saturation magnetization (Fig. 4(a-c)) which suggests that the ferromagnetic ordering is a thermodynamic first-order transition. Thermal hysteresis (~ 10 K) in magnetization measured by bulk magnetometry has also been observed in LPCMO films [27], though this value represents an average over the entire volume of the sample. The second remarkable feature about Fig. 4 is that when the saturation magnetization is normalized to the value at 20 K (reduced magnetization = M(T)/M(20 K)), the temperature dependencies of the reduced magnetizations are different for the different regions (Fig. 4 (d-e)). Near $T_{IM}$ or $T_{MI}$, only region II is noticeably magnetic, suggesting different ordering temperatures for different regions. Thus, we observed *coexistence of ferromagnetically ordered and magnetically disordered material as a function of depth*. The magnetic non-uniformity is seen for all temperatures below $T_{IM}$ and $T_{MI}$.

Previously, thermal hysteresis in magnetization measured by macroscopic techniques (e.g., SQUID and VSM) across MIT of LPCMO films and bulk polycrystals has been attributed to a difference in the dynamics of a magnetic phase [27, 33]. Coexistence of metastable functional domains and coupling to structural distortion can also be responsible for the hysteresis seen in each region of our LPCMO films. However, the presence of chemical non-uniformity is an additional complication that could couple to the magnetization, thereby influencing the signatures associated with a first order transition, namely, hysteresis and metastability.

In summary, we measured the depth dependence of the chemical and magnetic structures of an LPCMO film. The magnetic non-uniformity across the film's thickness was found to be related to its non-uniform chemical depth profile. EELS of the same sample suggests the presence of chemical non-uniformity (an increase in O/Mn ratio and concomitant enrichment of $Mn^{4+}$) at the surface and buried interface. XRR and PNR measurements also indicate a change of chemistry near the film's surface and buried interface. These regions have lower magnetization than the film bulk. The magnetization of the film bulk is uniform over length scales of nanometers. We showed that the thermal evolution of the saturation magnetizations for the surface and buried interface (regions I and III) are different than the film bulk (region II). Thus, different regions have different ordering temperatures. Further, we observed thermal hysteresis of the magnetization, which is indicative of a first order transition, and the magnitude of the hysteresis was different for the surface and buried interface compared to the film bulk. The chemical non-uniformity across the depth can lead to a modified effective coupling that can influence the ordering temperature and hysteresis. The variation of the depth dependent chemical, electronic and magnetic properties should be included in discussions of phase coexistence/separation in manganites. Further, the non-uniformity of chemical and magnetic properties should be considered in the interpretation of data acquired using characterization techniques that lack the ability to discriminate between different regions of the sample.

This work was supported by the Office of Basic Energy Science (BES), U.S. Department of Energy (DOE), BES-DMS funded by the DOE's Office of BES, the National Science Foundation (DMR-0804452) (HJ and AB), Materials Sciences and Engineering Division of the U.S. DOE (MV) and ERC starting Investigator Award, grant #239739 STEMOX (MAR). Los Alamos National Laboratory is operated by Los Alamos National Security LLC under DOE Contract DE-AC52-06NA25396. Research supported in part by ORNL's Shared Research Equipment (ShaRE) User Facility, which is sponsored by the Office of BES, U.S. DOE. We thank Masashi Watanabe (ORNL) for the Digital Micrograph PCA plug-in.

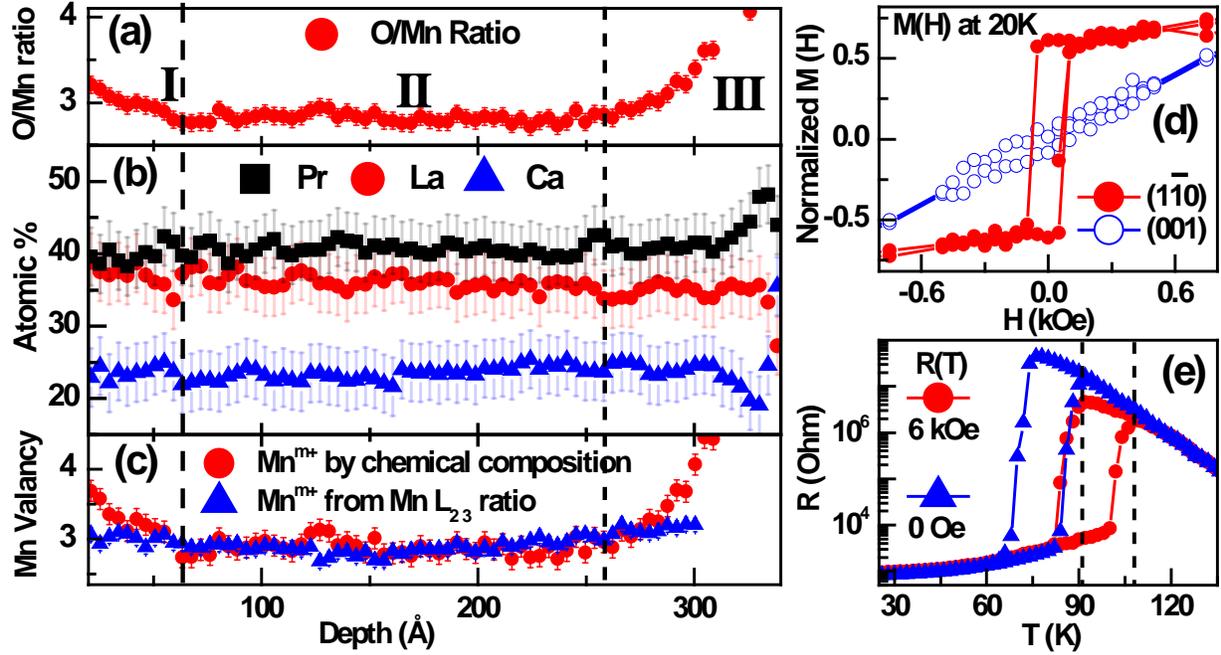

Fig. 1: (a)-(c) Depth sensitive EELS measurements: O/Mn ratio (a), atomic percentage (b) of different elements (La, Pr and Ca), and (c) Mn oxidation state calculated using both the chemical concentrations (●) and the Mn $L_{2,3}$ intensity ratio (▲) along the depth of the film (see text). (d): $M(H)$ hysteresis curve of the sample measured at $T$ = 20 K along two in-plane directions (1$\bar{1}$0) and (001) of NdGaO$_3$ substrate. (e): transport measurements of the film with and without magnetic field.

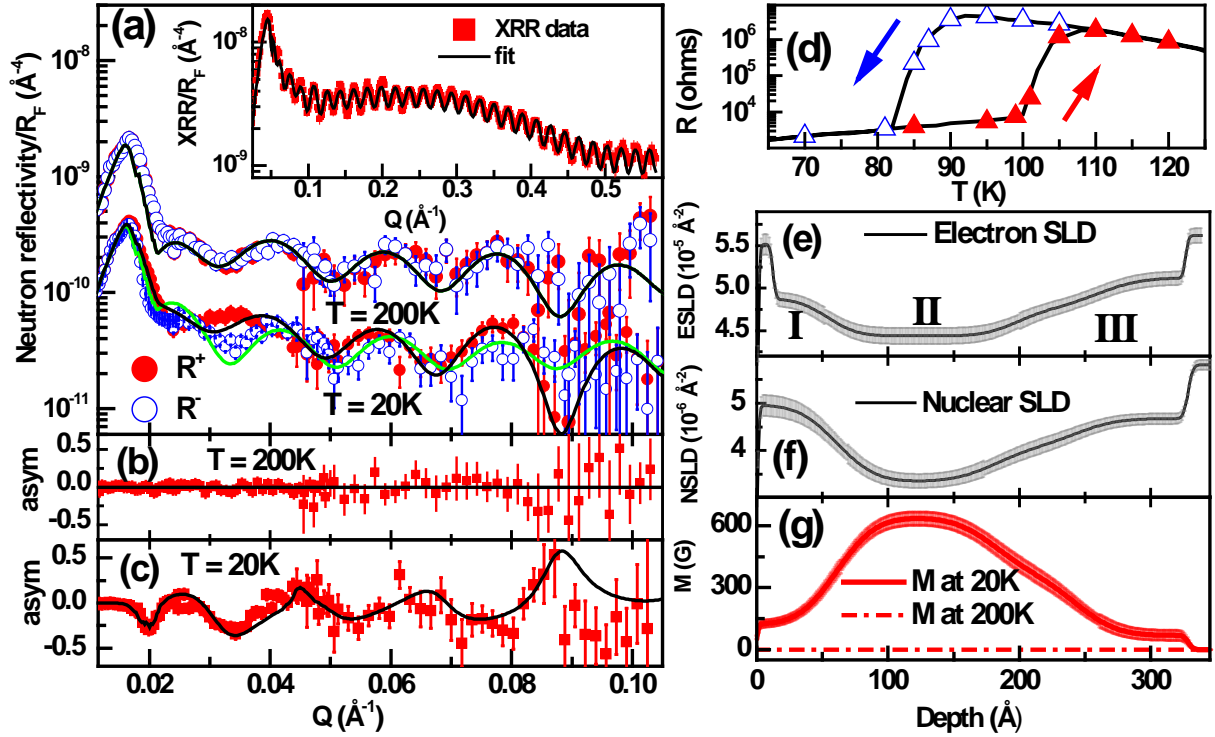

Fig. 2 (a): PNR Data from the sample at T = 200 K and 20 K. For clarity the PNR data at 200 K has been offset by 5. Inset of (a) show the XRR data from the film. $asym = (R^+ - R^-)/(R^+ + R^-)$, measurements at T = 200 K (b) and 20 K (c). The open and closed triangles on transport data (d) represents the temperature during cooling and warming where the PNR data were simultaneously acquired with the transport data. Fig. (e) shows the electron SLD (ESLD) depth profile which yields the solid curve in the inset of (a). Nuclear SLD (f) and magnetization (g) depth profile, which yields the solid curves in (a) at T = 200 K and 20 K.

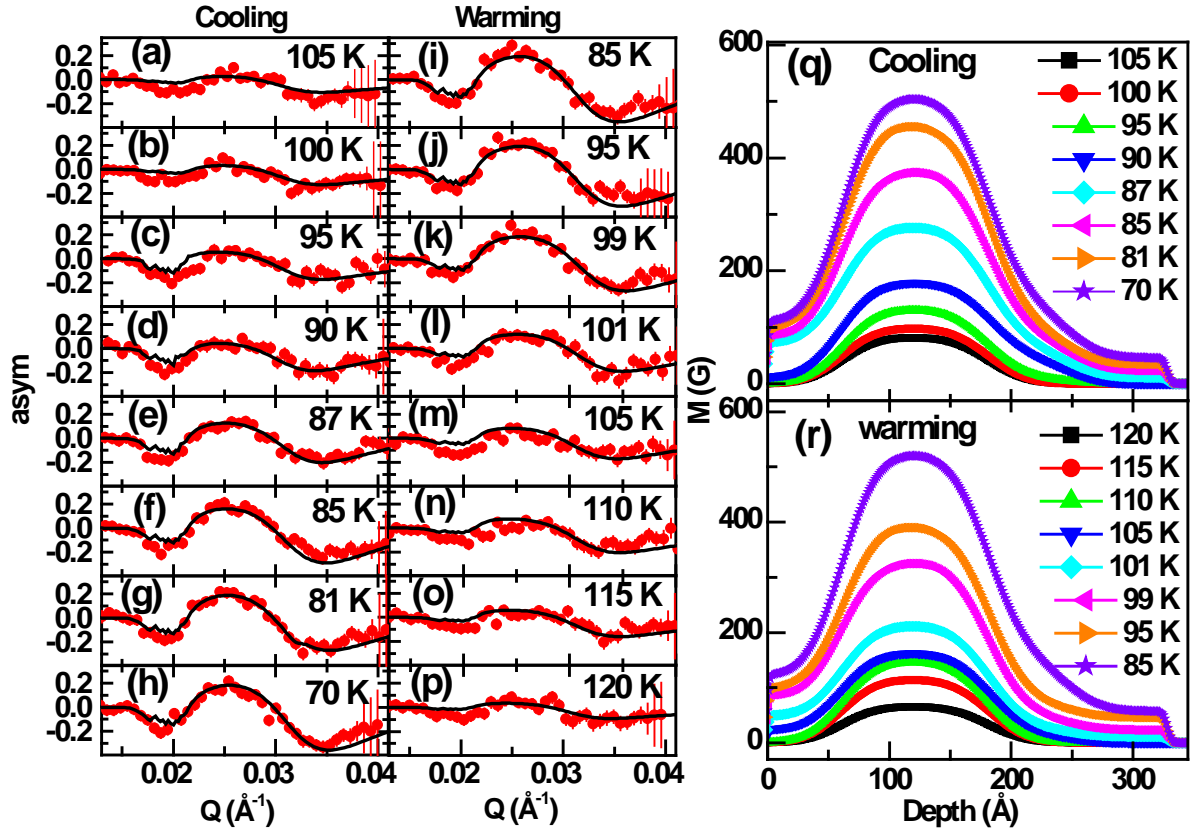

Fig. 3: Spin asymmetry (*asym*) during field cooling ((a) - (h)) and warming ((i)-(p)) across MIT. (q) - (r) show the magnetization (*M*) depth profile corresponding to (a)-(p).

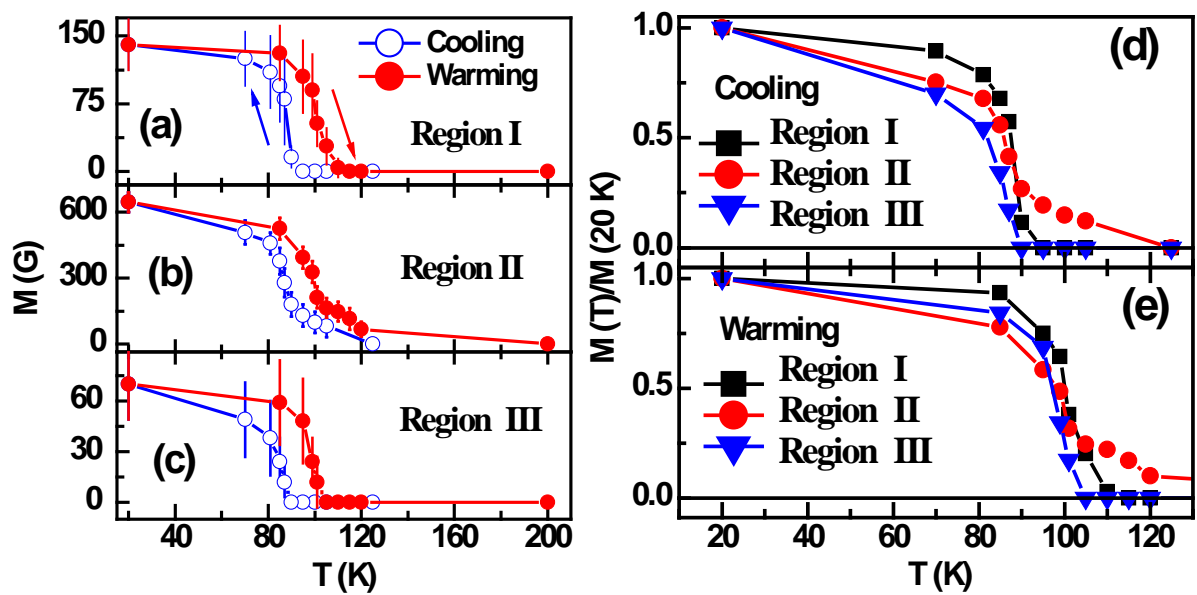

Fig. 4 (a)-(c): $M(T)$ curves for the different regions during cooling (blue) and warming (red). $M(T)$ normalized to $M(20\ \text{K})$ for the different regions during cooling (d) and warming (e).